\documentclass[prl,twocolumn,superscriptaddress,nobibnotes,letterpaper]{revtex4}

\usepackage[pdftex]{graphicx}
\usepackage[pdftex]{epsfig}
\usepackage{amsmath}
\usepackage{amssymb}
\usepackage{amsfonts}
\usepackage{color}
\usepackage{wrapfig}
\usepackage{eucal}
\usepackage{hhline}
\usepackage{threeparttable}
\usepackage{supertabular}
\usepackage{multirow}
\usepackage{tabularx}

\newcommand{\opd}[2]{\mbox{$\hat{#1}_\text{#2}^{\dagger}$}}
\newcommand{\op}[2]{\mbox{$\hat{#1}_\text{#2}$}}
\newcommand{\opdb}[2]{\mbox{$\hat{#1}_\text{#2}^{(\dagger)}$}}

\newcommand{\ket}[2]{\mbox{$\rvert{#1}\rangle_\text{#2}$}}

\def\be{\begin{equation}}
\def\ee{\end{equation}}
\def\bea{\begin{eqnarray}}
\def\eea{\end{eqnarray}}

\newcolumntype{Y}{>{\centering\arraybackslash}X}

\renewcommand{\figurename}{Figure}

\begin{document}

\pagenumbering{arabic}

\title{Non-classical correlations between single photons and phonons from a mechanical oscillator}\thanks{This work was published in Nature \textbf{530}, 313--316 (2016).}

\author{Ralf Riedinger}\thanks{These authors contributed equally to this work.}
\affiliation{Vienna Center for Quantum Science and Technology (VCQ), Faculty of Physics, University of Vienna, A-1090 Vienna, Austria}
\author{Sungkun Hong}\thanks{These authors contributed equally to this work.}
\affiliation{Vienna Center for Quantum Science and Technology (VCQ), Faculty of Physics, University of Vienna, A-1090 Vienna, Austria}
\author{Richard A.\ Norte}
\affiliation{Kavli Institute of Nanoscience, Delft University of Technology, 2628CJ Delft, The Netherlands}
\author{Joshua A.\ Slater}
\affiliation{Vienna Center for Quantum Science and Technology (VCQ), Faculty of Physics, University of Vienna, A-1090 Vienna, Austria}
\author{Juying Shang}
\affiliation{Photon Spot Inc., Monrovia, CA 91016, USA}
\author{Alexander G.\ Krause}
\affiliation{Kavli Institute of Nanoscience, Delft University of Technology, 2628CJ Delft, The Netherlands}
\author{Vikas Anant}
\affiliation{Photon Spot Inc., Monrovia, CA 91016, USA}
\author{Markus Aspelmeyer}
\email{markus.aspelmeyer@univie.ac.at}
\affiliation{Vienna Center for Quantum Science and Technology (VCQ), Faculty of Physics, University of Vienna, A-1090 Vienna, Austria}
\author{Simon Gr\"oblacher}
\email{s.groeblacher@tudelft.nl}
\affiliation{Kavli Institute of Nanoscience, Delft University of Technology, 2628CJ Delft, The Netherlands}


\begin{abstract}
Interfacing a single photon with another quantum system is a key capability in modern quantum information science. It allows quantum states of matter, such as spin states of atoms~\cite{Wilk2007,Stute2013}, atomic ensembles~\cite{Kuzmich2003,vanderWal2003} or solids~\cite{Yilmaz2010}, to be prepared and manipulated by photon counting and, in particular, to be distributed over long distances. Such light-matter interfaces have become crucial to
fundamental tests of quantum physics~\cite{Hensen2015} and realizations of quantum networks~\cite{Kimble2008}. Here we report non-classical correlations between single photons and phonons -- the quanta of mechanical motion -- from a nanomechanical resonator. We implement a full quantu
protocol involving initialization of the resonator in its quantum ground state of motion and subsequent generation and read-out of correlated photon-phonon pairs. The observed violation of a Cauchy-Schwarz inequality is clear evidence for the non-classical nature of the mechanical state generated. Our results demonstrate the availability of on-chip solid-state mechanical resonators as light-matter quantum interfaces. The performance we achieved will enable studies of macroscopic quantum phenomena~\cite{Romero-Isart2011} as well as applications in quantum communication~\cite{Stannigel2010}, as quantum memories~\cite{Chang2011} and as quantum transducers~\cite{Barzanjeh2012,Bochmann2013}.
\end{abstract}

\maketitle

\begin{figure*}[t]
\begin{center}
\includegraphics[width=1.9\columnwidth]{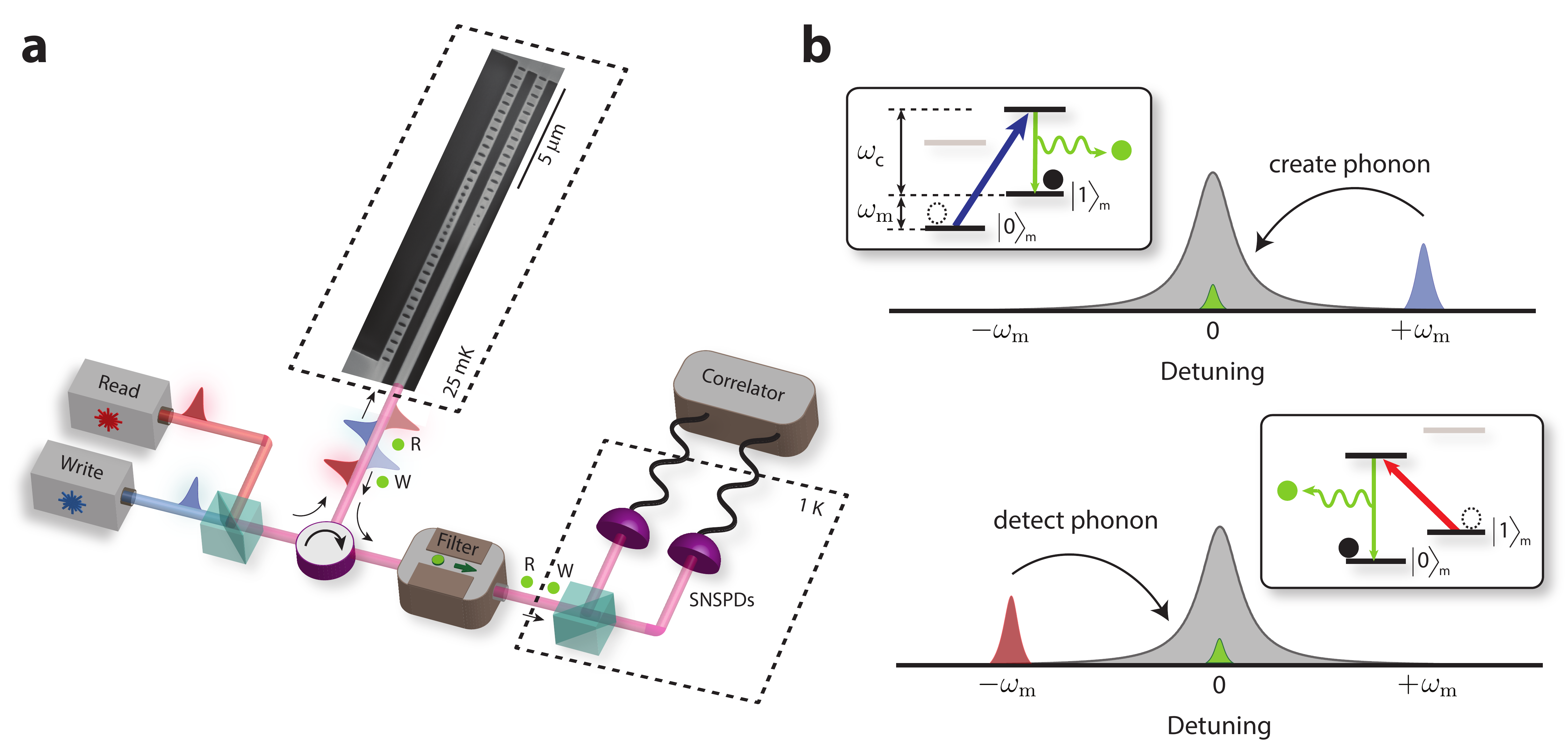}
\caption{\textbf{Generation and read-out of photon-phonon pairs.} \textbf{a}, Schematic of the experiment. Two independent lasers (stabilized to a wave-meter) are used to generate a sequence of 'write' and 'read' pulses with tunable time delay $\delta t$. They are sent through a circulator and drive a nanomechanical photonic crystal cavity (a scanning electron microscope image of which is shown in the inset) that is mounted inside	a dilution refrigerator at a base temperature of 25~mK, which prepares	the device in its quantum ground state of motion. For each pulse, Stokes and anti-Stokes Raman scattering creates single photons (green dots) from the write ($W$) and the read ($R$) pulse, respectively, that are emitted at a frequency $\omega_\textrm{c}$. The detuned pump fields are strongly suppressed by	optical filtering and only the Raman scattered photons are measured by two superconducting nanowire single-photon detectors (SNSPDs) in the output ports of a 50/50 beam-splitter. The time of each photon detection event is recorded and is then correlated in post-processing to obtain both auto- and cross-correlations of the emitted photons. A more detailed explanation of the experimental set-up is provided in Methods. \textbf{b}, Pulsed optomechanical interactions in frequency space. A blue-detuned write pulse realizes a two-mode squeezing interaction (blue and green pulses; see text). Cavity-enhanced Stokes Raman scattering generates a single phonon, stored as an excitation on the mechanical resonator, and a single ($W$) photon, which is emitted from the cavity on resonance (upper	panel). Reading out of the phonon utilizes a red-detuned read pulse, which swaps the mechanical excitation onto the optical cavity field, hence creating a single ($R$) photon (lower panel). The insets depict the relevant energy level diagrams for the two processes, reminiscent of the $\Lambda$-schemes in atomic Raman scattering. The grey bars indicate the energy levels that are not involved in the depicted process (Stokes or anti-Stokes), but in the other one.}
\label{fig:1}
\end{center} 
\end{figure*}

Over the past few years, nanomechanical devices have been discussed as possible building blocks for quantum information architectures~\cite{Stannigel2010,Wallquist2009}. Their unique feature is that they combine an engineerable solid-state platform on the nanoscale with the possibility to coherently interact with a variety of physical quantum systems including electronic or nuclear spins, single charges, and photons~\cite{Poot2012,Aspelmeyer2014}. This feature enables mechanics-based hybrid quantum systems that interconnect different, independent physical qubits through mechanical modes.

A successful implementation of such quantum transducers requires the ability to create and control quantum states of mechanical motion. The first step -- the initialization of micro- and nanomechanical systems in their quantum ground state of motion -- has been realized in various mechanical systems either through direct cryogenic cooling~\cite{OConnell2010,Meenehan2015} or laser cooling using microwave~\cite{Teufel2011b} and optical cavity fields~\cite{Chan2011}. Further progress in quantum state control has mainly been limited to the domain of electromechanical devices, in which mechanical motion couples to superconducting circuits in the form of qubits and microwave cavities~\cite{Aspelmeyer2014}. Recent achievements include single-phonon control of a micromechanical resonator by a superconducting flux qubit~\cite{OConnell2010}, the generation of quantum entanglement between quadratures of a microwave cavity field and micromechanical motion~\cite{Palomaki2013}, and the preparation of quantum squeezed micromechanical states~\cite{Wollman2015,Pirkkalainen2015,Lecocq2015}.

Interfacing mechanics with optical photons in the quantum regime is highly desirable because it adds important features such as the ability to transfer mechanical excitations over long distances~\cite{Stannigel2010,Safavi-Naeini2011a}. In addition, the available toolbox of single-photon generation and detection allows for remote quantum state control~\cite{Kimble2008}. However, micro- and nano-mechanical quantum control through single optical photons has not yet been demonstrated. One of the outstanding challenges is to achieve single-particle coupling rates that are sufficiently large to alleviate
effects of optical and mechanical decoherence in the system, that is, single-photon strong co-operativity. Some of the largest optomechanical couplings have been reported in nanomechanical photonic crystal cavities~\cite{Safavi-Naeini2010}, but are still two orders of magnitude short of that regime.
Although low coupling rates can be overcome in principle by a strong and detuned coherent drive field~\cite{Aspelmeyer2014}, such measures typically result in unwanted heating of the mechanical device (see Methods).

Here we take a different approach that allows us to circumvent these problems and to realize quantum control of single phonons through single optical photons. We use a probabilistic scheme based on the well-known DLCZ protocol (Duan, Lukin, Cirac and Zoller)~\cite{Duan2001}, which, in its original form, uses Raman scattering for efficient generation and read-out of collective spin states of atomic ensembles. In essence, the scheme generates entanglement through single-photon interference and post-selection, which does not require strong coupling~\cite{Cabrillo1998}. In the context of mechanical quanta, this protocol has been used in an experiment to entangle high-frequency (40~THz) optical phonons of two bulk diamond lattices~\cite{Lee2011}. However, the small interaction and coherence times of such phonons are incompatible with their use in quantum transduction
and storage, and so it is necessary to take this approach to the level of chip-scale optomechanical systems. In addition, we minimize absorption heating by using short optical pulses in a cryogenic environment~\cite{Meenehan2015}. The combination of these techniques allows us to overcome the previous limitations and realize a photon-phonon quantum interface.

Our experiment complements previous work on single- and two-mode (opto-)mechanical squeezing in microwave circuits~\cite{Palomaki2013,Wollman2015,Pirkkalainen2015,Lecocq2015}. Although these experiments were based on the same underlying interactions, they involved homodyne or heterodyne detection of light to access continuous-variable degrees of freedom of a quantum state -- specifically, quadrature fluctuations in the mechanical and optical canonical variables. In contrast, the DLCZ scheme uses photon counting, which allows access to discrete quantum variables -- here, in form of energy eigenstates (phonons) of the mechanical motion -- and thereby enables realistic architectures for entanglement distribution and quantum networking~\cite{Kimble2008}.

The mechanical system studied here is a micro-fabricated silicon photonic crystal nanobeam structure (Fig.~\ref{fig:1}a). Such optomechanical crystals co-localize optical and mechanical modes and couple them via a combination of radiation pressure and photostriction~\cite{Aspelmeyer2014}. Our device exhibits an optical cavity resonance at wavelength $\lambda_\textrm{c}$ = 1,556~nm and a mechanical breathing mode at frequency $\omega_\textrm{m}/2\pi$ = 5.3~GHz. The cavity decay rate (full-width at half-maximum, FWHM) is $\kappa_\textrm{c}/2\pi$ = 1.3~GHz and the mechanical quality factor at cryogenic temperature is $Q_\textrm{m} = 1.1\cdot10^6$ (see Methods). Pulsed optical driving at laser frequency $\omega_\text{L}=\omega_\text{c}\pm\omega_\text{m}$ (in which $\omega_\text{c}=2\pi c/\lambda_\textrm{c}$ is the cavity frequency and $c$ is the vacuum speed of light) allows to realize two different types of interactions on the basis of cavity-enhanced Stokes ($+$) and anti-Stokes ($-$) Raman scattering (Fig.~\ref{fig:1}b). A blue-detuned pulse ($\omega_\text{L}=\omega_\text{c}+\omega_\text{m}$) results in two-mode squeezing with interaction Hamiltonian
$H_\text{tms}\propto\hbar g_\text{0}\left(\opd{a}{m}\opd{a}{o}+\op{a}{m}\op{a}{o}\right)$, in which $\opdb{a}{m}$ and $\opdb{a}{o}$ are the creation (annihilation) operators of the mechanical and optical mode, respectively, $g_\text{0}$ is the effective optomechanical coupling rate (here, $g_\text{0}/2\pi=825$~kHz; see Methods) and $\hbar$ is the reduced Planck constant. This interaction generates photon-phonon pairs in close analogy to the photon-photon pairs generated in parametric down-conversion~\cite{Wu1986}. A red-detuned pulse ($\omega_\text{L}=\omega_\text{c}-\omega_\text{m}$) allows read-out of the mechanical state through the optomechanical beam-splitter interaction $H_\text{bs}\propto\hbar g_\text{0}\left(\op{a}{m}\opd{a}{o}+\opd{a}{m}\op{a}{o}\right)$, in which an anti-Stokes scattering event realizes a state swap between the mechanical and optical cavity mode.

Our protocol consists of three distinctive steps. First, we initialize the mechanical system in its quantum ground state of motion by cryogenic cooling. Second, a short blue pulse creates a photon-phonon pair and leaves the originally empty mechanical and optical modes $\ket{0}{m}$ and $\ket{0}{o}$ at frequencies $\omega_\text{m}$ and $\omega_\text{c}$, respectively, in the state $\ket{\Phi}{om}=\ket{00}{}+\sqrt{p}\ket{11}{}+p\ket{22}{}+\mathcal{O}(p^{3/2})$. Here $p$ is the probability for a single Stokes scattering event to take place. Residual heating through optical absorption introduces additional noise to the state (see Methods). Finally, a strong red pulse is used to read out the phonon state via emission of an anti-Stokes scattered photon~\cite{Cohen2015}. We confirm the non-classical photon-phonon correlations on the basis of an observed violation of a Cauchy-Schwarz inequality for the cross-correlation of the coincidence measurements between the Stokes and anti-Stokes photons~\cite{Kuzmich2003}.

\begin{figure}[ht!]
\includegraphics[width=.95\columnwidth]{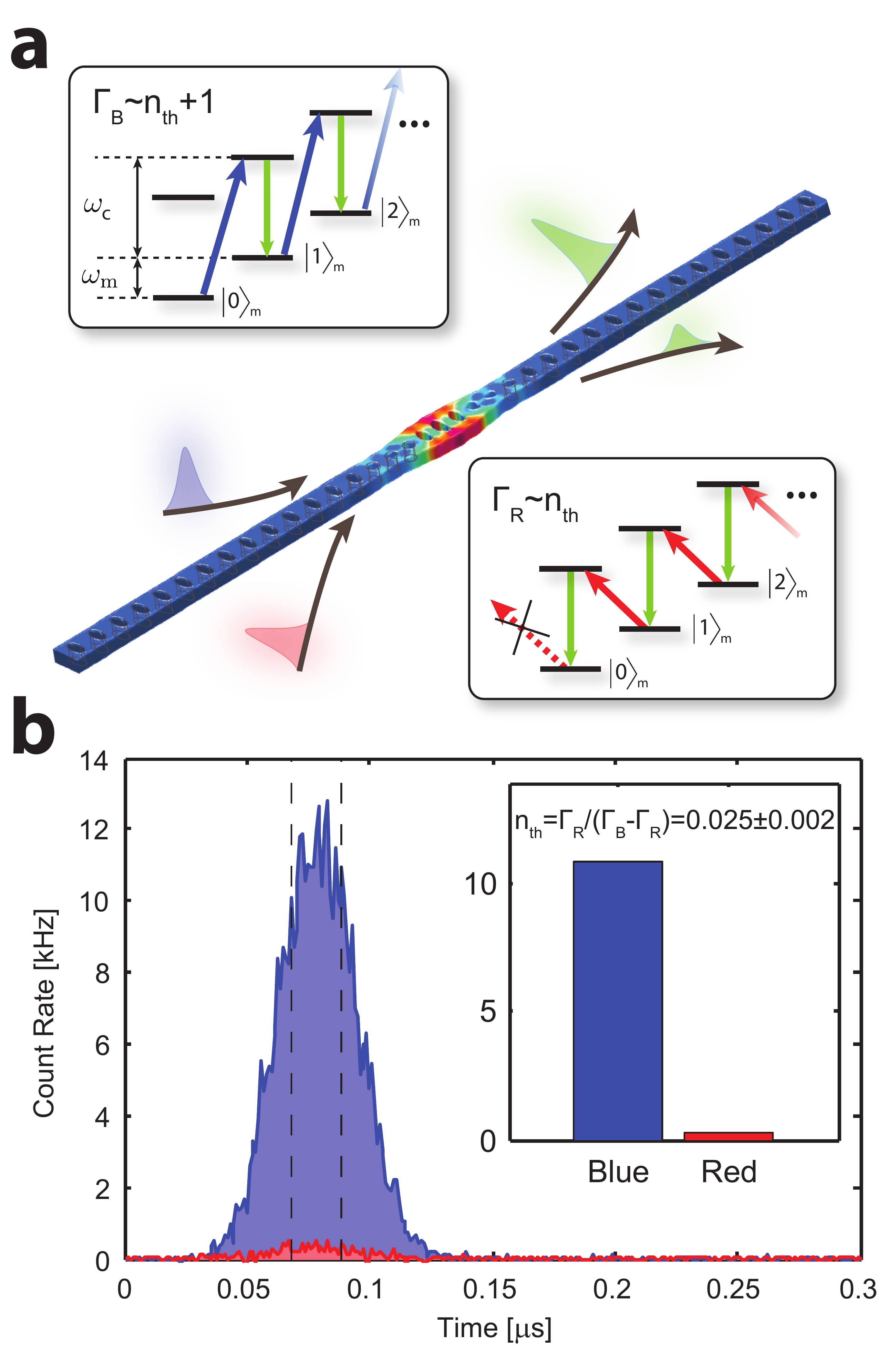}
\caption{\textbf{Mechanical quantum ground state preparation.} \textbf{a}, Principle of sideband thermometry. The finite element method simulation depicted in the main panel shows the structure of the mechanical breathing mode under investigation. The upper (lower) inset shows the energy level scheme in case of blue- (red-) detuned pumping and the resultant cavity-enhanced Stokes (anti-Stokes) scattering. The corresponding scattering rates $\Gamma_\text{R}$ and $\Gamma_\text{B}$ are proportional to thermal occupation of the mechanics	$n_\text{th}$ and $n_\text{th}+1$, respectively, and hence show a strong asymmetry when the mechanics are close to the quantum ground state. \textbf{b}, Sideband asymmetry. The optomechanical device is pumped with a sequence of alternating blue- and red-detuned optical pulses at frequency $\omega_\text{c}\pm\omega_\text{m}$ (optical energy per pulse $E_\text{opt} = 33$~fJ; FWHM of 28.4~ns; 500~$\mu$s separation of pulse sequences). Shown are the count rates recorded by the SNSPDs as a function of the arrival time of the scattered photons (blue, blue-detuned pulse; red, red-detuned pulse). This data has been corrected for leakage of pump photons through the optical filters, which was independently measured and subtracted from our data (see Methods). The inset shows a histogram of the total counts that are obtained when averaging over a 20~ns window centred on the peak (within the dashed lines). The pronounced asymmetry in the rates (of more than a factor of 40) corresponds to a thermal occupancy of $n_\text{th} = \Gamma_\text{R}/(\Gamma_\text{B}-\Gamma_\text{R}) = 0.025\pm0.002$ and to a mode temperature of 69~mK.}
\label{fig:2}
\end{figure}

Precooling of the nanomechanical device is performed using a dilution refrigerator that operates at a base temperature of approximately 25~mK. If the mechanical system is in its quantum ground state of motion, then anti-Stokes processes cannot occur because no additional phonons can be extracted to support the scattering. This is in contrast to Stokes processes, which deposit mechanical energy and hence can always occur. As a consequence, the asymmetry in the scattering rates of these two processes is a direct measurement of the mean thermal phonon occupancy $n_\text{th}$. Using such photon-counting based sideband thermometry~\cite{Meenehan2015}, we find $n_\text{th}\lesssim0.025$ (see Fig.~\ref{fig:2}).

\begin{figure*}[ht!]
\includegraphics[width=1.95\columnwidth]{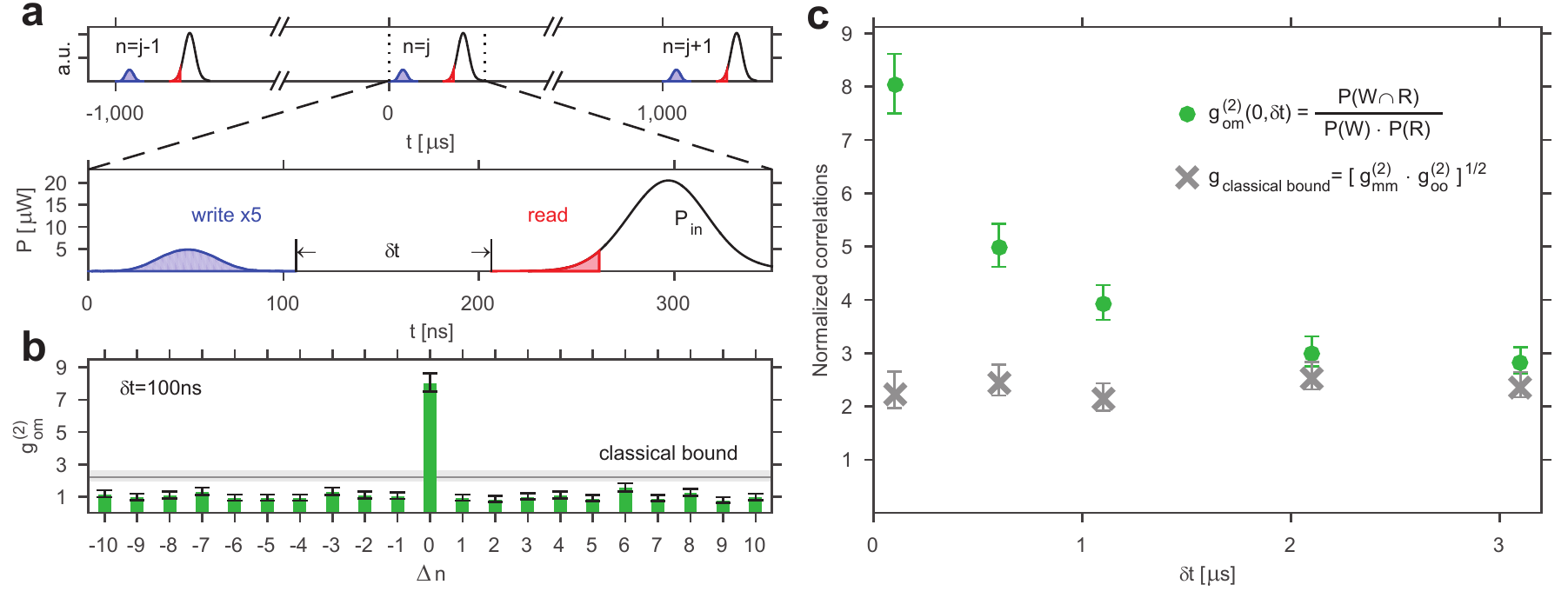}
\caption{\textbf{Non-classical photon-phonon correlations.} \textbf{a}, Driving pulse sequence. A pair of one write (blue) and one read (red) pulse is sent to the device every 1~ms. The long idle phase between pulse pairs ensures the ground-state initialization by cryogenic cooling. Each pulse sequence is labelled with a number ($n$). The read pulse is delayed by $\delta t$ with respect to the write pulse, and only the first 55~ns, equivalent to a read-pulse power of about 50~fJ, are used for the data evaluation. This reduces the influence of absorption heating while maintaining reasonable state swap fidelity. \textbf{b}, Violating a Cauchy-Schwarz inequality. Shown is the cross-correlation (green bars) between the mechanical (read pulse) and optical state (write pulse) for $\delta t=100$~ns, as well as the classical (Cauchy-Schwarz) bound obtained from the autocorrelations at $\Delta n=0$ (grey horizontal line, shading indicates a 68\% confidence interval; see text). For photon-phonon pairs that emerge from different pulse sequences ($\Delta n\neq0$) the Cauchy-Schwarz inequality is fulfilled, $\langle g_\text{om}^{(2)}(\Delta n\neq0,100~\text{ns})\rangle=1.04\pm0.04$, consistent with statistical independence. For pulses from the same pair, the cross-correlation $g_\text{om}^{(2)}(0,100~\text{ns})$ clearly exceeds the classical bound. $g_\text{om}^{(2)}$ can be interpreted as the ratio of heralded phonons $n_\text{h}$ to unheralded (thermal) phonons $n_\text{th}$ at the time of the read pulse. \textbf{c}, Storage of non-classical correlations. Shown is the dependence of the cross-correlation on the time delay $\delta t$ between the write and read pulses. For each data point, the classical bound is measured independently through the normalized autocorrelation functions of the write ($W$) and read ($R$) photons. For increasing $\delta t$, the photon-phonon cross-correlations decrease, but stay above the classical limit even beyond 1~$\mu$s. The main contribution to the loss of correlation is heating by absorption of the write pulse (see Methods). All error bars represent a 68\% confidence interval.} 
\label{fig:3}
\end{figure*}

We create the desired photon-phonon pairs using a blue-detuned 'write' pulse that is sufficiently weak to minimize the effects of residual absorption heating (FWHM, 28.4~ns; energy, 40~fJ). We find the relevant probability to generate a Stokes scattered photon on cavity resonance to be $p\approx3.0$\%. Subsequently, a red-detuned 'read' pulse (effective length, 55~ns; energy of approximately 50~fJ) is injected at a time delay $\delta t$ (see Fig.~\ref{fig:3}a), resulting in a phonon-to-photon conversion efficiency of approximately 3.7\% (see Methods).

We correlate the measured Stokes- and anti-Stokes photons via the cross-correlation function $g_\text{om}^{(2)}(\Delta n, \delta t) = P(W\cap R)/\left[P(R)P(W)\right]$, which is computed for read and write pulses originating from pulse sequences from different trials separated by $\Delta n$ iterations (see Fig.~\ref{fig:3}). $P(W\cap R)$ is the probability for a joint detection of both a Stokes ($W$, 'write') and an anti-Stokes ($R$, 'read') photon from these pulses, and $P(W)$ and $P(R)$ are the unconditional probabilities to detect either of the two photons. For all pair correlations of classical origin, the value of $g_\text{om}^{(2)}$ is bounded by a Cauchy-Schwarz inequality of the form~\cite{Kuzmich2003} $g_\text{om}^{(2)}(0, \delta t)\leq\left[g_{\text{oo,}\delta t}^{(2)}(0)g_{\text{mm,}\delta t}^{(2)}(0)\right]^{1/2}$, in which $g_{\text{oo,}\delta t}^{(2)}(0)$ and $g_{\text{mm,}\delta t}^{(2)}(0)$ are the autocorrelation functions for the optical and mechanical mode, respectively (see Methods). A violation of this inequality~\cite{Kuzmich2003,Clauser1974,Foertsch2013} is an unambiguous measure for the non-classicality of the generated photon-phonon state. The Cauchy-Schwarz inequality for coincidence detection marks a well-defined border between the quantum and classical domain. It is based on the fact that the Glauber-Sudarshan phase-space function, or P-function, is positive definite for every classical field. This places a funda­mental limit on the relative strength of measurable cross-correlations versus autocorrelations between classical fields. Previous applications of this limit include the distinction between the classical and quantum field theoretical predictions for the photoelectric effect~\cite{Clauser1974}, and the storage and retrieval of non-classical states in the collective emission from an atomic ensemble~\cite{Kuzmich2003}. A detailed derivation of the Cauchy-Schwarz inequality for the case of non-stationary fields, as are being used here, is provided in ref.~\cite{Kuzmich2003}.

We find a clear violation for an extended regime of time delays. Figure~\ref{fig:3}b shows the value of $g_\text{om}$ at a time delay of 100~ns. For pairs emitted from the same pulse sequence ($\Delta n=0$) we find that $\left\lbrace g_\text{om}^{(2)}(0,100~\text{ns}) = 8.0+0.6-0.5\right\rbrace\nleqslant\left\lbrace\left[g_\text{oo,100~ns}^{(2)}(0) g_\text{mm,100~ns}^{(2)}(0)\right]^{1/2} = 2.09+0.23-0.16\right\rbrace$, which obviously violates the classical bound. As expected, pairs emitted from different pulse sequences ($\Delta n\neq0$) are uncorrelated and hence fulfil the inequality. Upon increasing the time delay further, we find a violation even beyond $\delta t=1~\mu$s (see Fig.~\ref{fig:3}c), which demonstrates that we can store and retrieve non-classical states for an extended time interval. Nevertheless, the lifetime of these non-classical correlations is still much shorter than the lifetime of the mechanical excitations, $Q/\omega_\text{m}\approx34~\mu$s. We attribute this to the fact that the dynamics are dominated by heating caused by absorption of pump photons, which after some onset time drives the mechanical system towards a thermal state (see Methods). As a consequence, reducing the energy of the write pulse further should allow non-classical correlations to be maintained for much longer times. In addition, upon further reduction of the absorption heating of the read pulse, even higher values for the cross-correlation are obtained.

The cross-correlation is also linked to the autocorrelation of the heralded mechanical state. If one considers two-mode optomechanical squeezing acting on an initial mechanical thermal state, and if $g_\text{om}^{(2)}\gg1$ -- as is the case in our experiment -- then one obtains $g_\text{mm,heralded}^{(2)}\approx4/\left(g_\text{om}^{(2)}-1\right)$. The largest value for $g_\text{om}^{(2)}$ observed in our experiment was $g_\text{om}^{(2)}(0,100~\text{ns})= 19.6-2.8+3.9$ (using an energy of 1.7~fJ in the first 30~ns of the read pulse; see Methods). In other words, our system should allow for a Hanbury Brown and Twiss experiment with phonons yielding $g_\text{mm,heralded}^{(2)}\approx0.22$. A direct measurement of this value with the current experimental parameters is difficult without a prohibitively large number of pulse sequences.

In summary, we have demonstrated non-classical correlations between single photons and phonons from a nanomechanical resonator. This is a crucial step towards on-chip photon-phonon quantum interfaces, which are relevant for future solid-state based quantum information and communication architectures. For example, the observed photon-phonon correlation of $g_\text{om}^{(2)}=19.6$ suggests that conditional mechanical Fock-state preparation should be possible with fidelities exceeding 85\% (see Methods). The ability to store and retrieve non-classical states over extended storage times that we reported also shows that nano-optomechanical resonators are a promising candidate for quantum memories. The performance of the system we have
demonstrated constitutes an improvement of almost two orders of magnitude on previous lifetimes of stored non-classical single-phonon states~\cite{OConnell2010}. Finally, photon-phonon conversion on the single particle level is required to extend the ongoing efforts on mechanically transduced conversion between microwave and optical fields~\cite{Bochmann2013} into the quantum domain~\cite{Barzanjeh2012}.

\def\urlprefix{}

\textbf{Acknowledgments}\ Acknowledgements We thank K.\ Hammerer and S.\ Hofer for discussions, and T.\ Graziosi, J.\ Hill, J.\ Hoelscher-Obermaier, Y.\ Liu, L.\ Procopio, A.\ Safavi-Naeini, E.\ Schafler, G.\ Steele and W.\ Wieczorek for experimental support. We acknowledge assistance from the Kavli Nanolab Delft, in particular from M.\ Zuiddam and F.\ Dirne. This project was supported by the European Commission (cQOM, SIQS, IQUOEMS), a Foundation for Fundamental Research on Matter (FOM) Projectruimte grant (15PR3210), the Vienna Science and Technology Fund WWTF (ICT12-049), the European Research Council (ERC CoG QLev4G), and the Austrian Science Fund (FWF) under projects F40 (SFB FOQUS) and P28172. R.\ R.\ is supported by the FWF under project W1210 (CoQuS) and is a recipient of a DOC fellowship of the Austrian Academy of Sciences at the University of Vienna.


\clearpage

\renewcommand{\figurename}{Methods Figure}

\section{Methods}

\setcounter{figure}{0}

\subsection{Device fabrication and characterization}

\begin{figure}[ht!]
\includegraphics[width=.95\columnwidth]{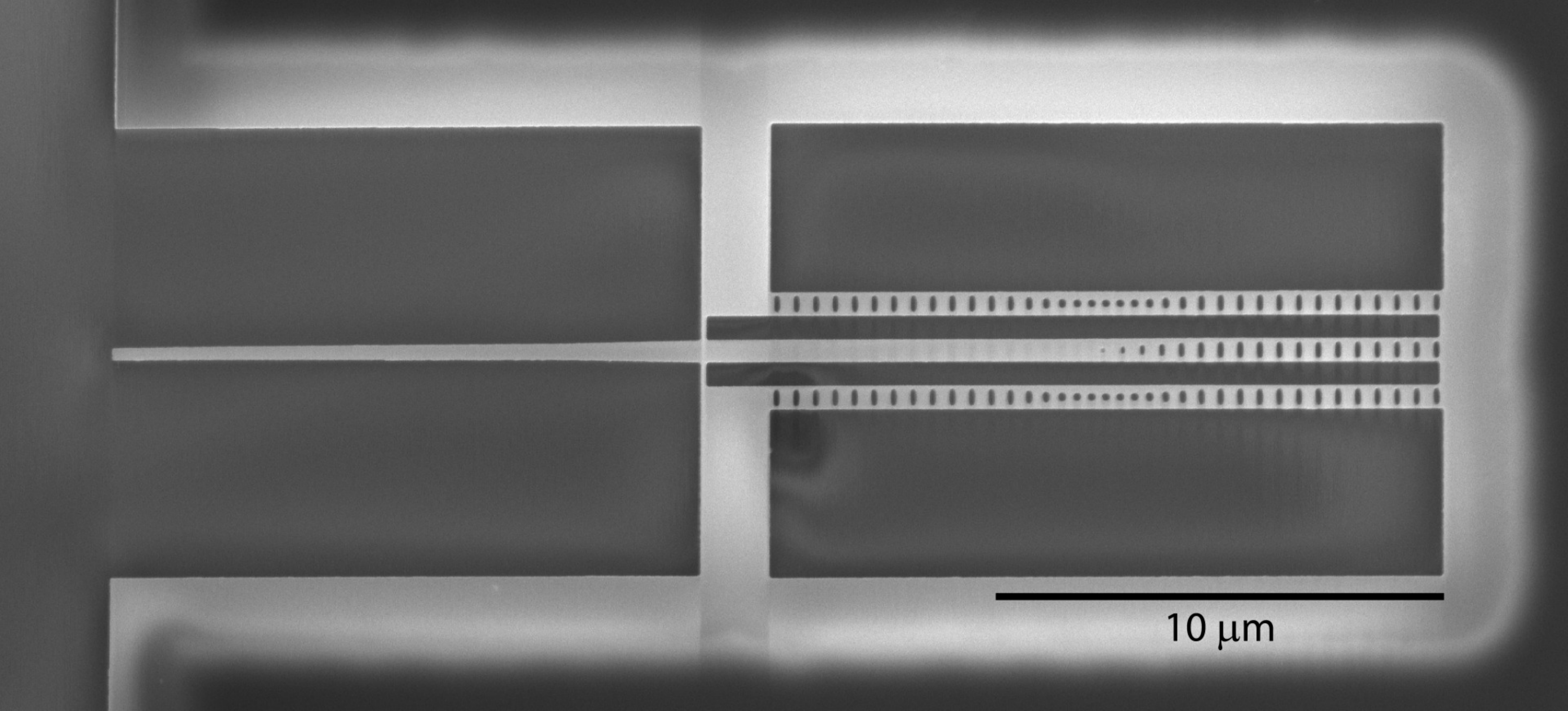}
\caption{\textbf{Optomechanical device.} Shown is a scanning electron microscope image of a set of nanobeams, which are fabricated in silicon, as described in the text. Light is coupled into the central, adiabatically tapered waveguide through a lensed optical fiber (not shown) from the left of the image. The field then evanescently couples to each nanobeam (top and bottom). The two devices have slightly different resonance frequency, which makes it possible to distinguish them.}
\label{fig:M1}
\end{figure}

\begin{figure*}[ht!]
\includegraphics[width=1.8\columnwidth]{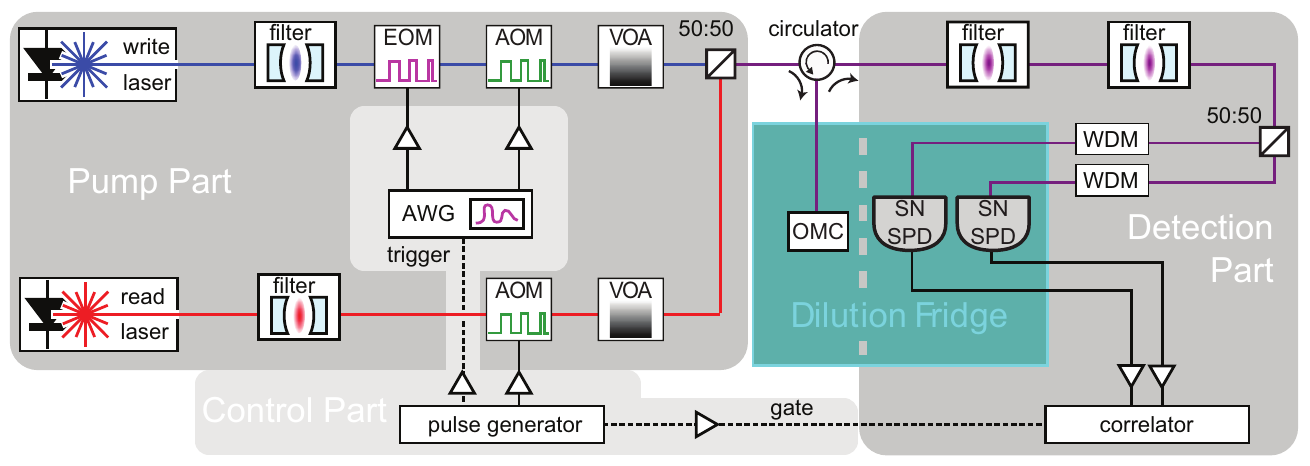}
\caption{\textbf{Detailed experimental setup.} See the Methods text for a detailed description.}
\label{fig:M2}
\end{figure*}

The optomechanical device used for this experiment (see Methods Figure~\ref{fig:M1}) is fabricated from a silicon-on-insulator wafer, with a device layer thickness of 250~nm and 3~$\mu$m of buried oxide. The structures are patterned using an electron beam writer and are then transferred into the top silicon layer in an SF$_6/$O$_2$ atmosphere using a reactive ion-etcher. The devices are finally released and undercut using concentrated hydrofluoric acid. We design the nanobeams such that the fundamental mechanical breathing mode is at 5.3~GHz (cf.\ Figure~\ref{fig:2}a, main text) and the optical resonance is around 1550~nm (the measured wavelength for the device used here is 1556~nm)~\cite{Chan2012}. The optical and the mechanical modes are co-localized in the center of the beam, where we create a defect region of the photonic- and phononic-bandgap, allowing for an optomechanical coupling rate $g_0/2\pi=825$~kHz. In order to minimize the thermalization time to the surrounding bath we opted, unlike previous designs, to not use any additional phononic shielding. As a consequence, the mechanical quality factors at base temperature are found to be around $1.1\cdot10^6$ (see section \textit{Mechanical response to optical pulses}), compared to values above $10^7$ with a phononic shield~\cite{Meenehan2015}. The laser pulses are coupled directly into a tapered waveguide through an optical fiber with a lensed tip~\cite{Meenehan2014}, achieving efficencies of about 60\%. The optical mode of the nanobeam is evanescently coupled to the waveguide, which is terminated with a periodic array of holes, acting as a mirror, allowing us to collect the light in reflection. For this experiment we chose a critically coupled device (internal losses equal external losses) with an optical linewidth $\kappa_\text{c}/2\pi$ of approximately 1.3~GHz. This places us well within the so-called resolved-sideband regime ($\omega_\text{m} > \kappa_\text{c}$).

\subsection{Setup}

In this section, we provide a detailed description of the experimental setup. It consists of a 'pump part', 'detection part', and the 'electronic control part' (cf.\ Methods Figure~\ref{fig:M2}).\\

\textit{Pump part}. We use two identical, tunable continuous-wave (CW) lasers (New Focus 6728) as our light sources. The lasers are detuned and stabilized to the blue and red side respectively of the device’s cavity resonance (1556.21~nm). The detuning is set to be the mechanical frequency (5.307~GHz). The two lasers separately pass through voltage-controlled tunable optical filters (MicronOptics FFP-TF2, free spectral range $\sim$18~GHz, bandwidth $\sim$50~MHz) to suppress any potential background emissions dispersed in frequency space. In order to create short optical pulses we modulate the filtered CW fields using acousto-optic modulators (AOM; IntraAction) and an additional electro-optic amplitude modulator (EOM; EOSpace). We employ variable optical attenuators (VOA; Sercalo) on each path to control the pulse power. The pulses are combined on a variable optical coupler and then sent to the device in the dilution refrigerator (Vericold E21) via an optical circulator. At the device (OMC; optomechanical crystal), the optomechanical interaction with the blue (red) detuned pulses generates down-(up-) converted photons, whose frequency is on resonance with the device's optical cavity frequency. The scattered photons are reflected back from the OMC into the optical fiber and routed to the detection part through the output port of the circulator.\\

\textit{Detection part}. Two voltage-controlled optical filters (MicronOptics FFP-TF2, specification as above) are installed in series at the beginning of the detection path. These filters are tuned on resonance with the OMC cavity frequency such that they only allow (anti-) Stokes scattered photons to be transmitted, while strong off-resonant pump photons are rejected (suppression  of about 84~dB). After the filters, a 50:50 beam splitter divides the path. Each output is additionally filtered by broadband wavelength-division multiplexors (WDM), and fiber-coupled to two superconducting nanowire single photon detectors (SNSPD; PhotonSpot, detection efficiency $\sim$90\%, dark count rate $<$10~Hz). The SNSPDs are mounted on the 1~K plate inside the dilution refrigerator. Upon receiving a photon the SNSPD generates a brief voltage spike, which is then electrically registered by a time-correlated single photon counting module (TCSPC; PicoQuant TimeHarp 260 NANO). The overall efficiency of detecting a photon leaving the OMC is $\sim$2.7\% (see below).\\

\textit{Control part}. In order to generate programmable optical pulses and to detect photons synchronously, we use a digital pulse generator (DPG; Highland Technology P400) and an arbitrary waveform generator (AWG; Agilent Technologies 81180A). We first program the DPG to generate a TTL gate voltage signal for the AOM on the read (red) path and to trigger the TCSPC synchronously. The DPG additionally triggers the AWG, which then generates a TTL gate voltage for the AOM and a voltage pulse for the EOM on the write (blue) path.

\begin{figure*}[ht!]
\includegraphics[width=1.95\columnwidth]{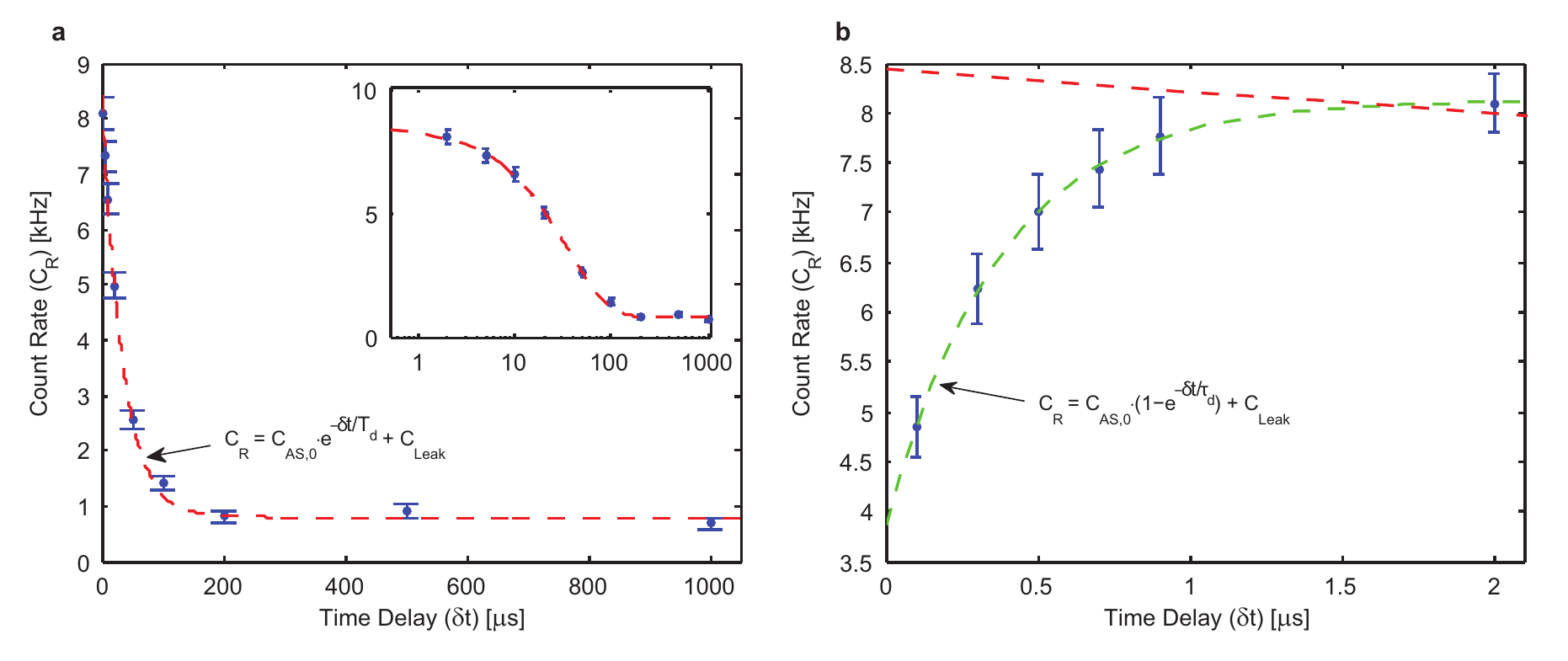}
\caption{\textbf{Pump-probe measurement of the mechanical response.} We send in a brief, intense blue detuned optical pulse (pump) and measure the mechanical response via red detuned optical probe pulse as a function of pump-probe time delay ($\delta t$). \textbf{a} Long-term mechanical response. The result fits well with a simple exponential decay (red dashed line; see the equation in the plot) with a damping time constant ($T_\text{d}$) of 34.4~$\mu$s. The inset shows the same data/fit with a logarithmic scale on the $x$ axis. $C_\text{AS,0}$ is the extrapolated $C_\text{AS}(\delta t=0)$. \textbf{b} Short-term mechanical response. The data is fitted to a simple exponential curve (green dashed line; see the equation in the plot). The fitted time constant ($\tau_\text{d}$) is 0.37~$\mu$s. The fit results of long-term response (red dashed line) projected to 0~$\mu$s delay is also shown for comparison. As the pump pulse had 5 times stronger energies than the write pulses in the correlation experiment, it is expected that the delayed heating occurs on longer time scale, due to the temperature dependence of the thermal conductivity of silicon~\cite{Meenehan2015}. Error bars in \textbf{a} and \textbf{b} represent a 68\% confidence interval.}
\label{fig:M3}
\end{figure*} 

\subsection{Mechanical response to optical pulses}

Over the past few years, several experiments have demonstrated precise control over optical and mechanical states through continuous optomechanical driving, including coherent state transfer~\cite{Palomaki2013,Fiore2011,Verhagen2012} and microwave-to-optics conversion~\cite{Bochmann2013,Andrews2014,Bagci2014}. Due to the unavailability of the regime of single-photon strong cooperativity, strong drive fields have to be used in order to achieve the wanted coupling strength~\cite{Akram2010}. This leads to unwanted heating effects, in particular in the optical domain. Since the mechanism of optical absorption couples only indirectly to the mechanical mode of interest~\cite{Meenehan2014}, using short optical pulses as nonstationary drive fields can substantially suppress the heating on short time scales -- in particular at low temperatures~\cite{Meenehan2015}.

Here, we probe the thermal response of the mechanical mode by pump-probe type measurements; we first send a short blue-detuned pump pulse onto the OMC cavity to intentionally heat the mode, and subsequently inject a red-detuned probe pulse to read out the mode’s phonon occupancy. By repeating the experiment with varying time delay between the pump and the probe pulses, we monitor time-dependent evolution of the mode’s phonon occupancy with a fixed initial impulse heating. The time delay $\delta t$ is defined as the delay between the end of the pump (blue) pulse and the start of the probe (red) pulse detection window, as indicated in Figure~\ref{fig:3}a in the main text. In that way the probing is performed after the optical absorption of the pump photons is completed. In order to ensure that the mechanical mode fully re-thermalizes to the bath, we set the duty cycle of sending another blue pulse after the red pulse to be one millisecond. For an improved signal to noise ratio, the pulse energies of the blue (200~fJ) and red pulses (2~pJ) used here are substantially larger than in the cross-correlation measurements.

The effective mode temperature is inferred from the average count rate observed after sending the red pulse ($C_\text{R}$). $C_\text{R}$ can be decomposed into three terms:\ (1) the rate proportional to the (on-resonance) anti-Stokes Raman scattered pump photons ($C_\text{AS}$), (2) the term corresponding to pump photons leaked through the optical filters ($C_\text{Leak}$), and (3) the additional anti-Stokes scattering term due to heating (ref.~\cite{Meenehan2015}) of the mode during the readout pulse ($C_\text{Heat}$). We minimize $C_\text{Heat}$ by only taking into account the first 30~ns of the red pulse as 'logical' red pulse. $C_\text{Leak}$ gives a constant offset to the signal. $C_\text{AS}$ directly reflects the mode's effective temperature, as the anti-Stokes scattering rate is proportional to the average number of phonons ($n_\text{m}$) in the mode (see Figure~\ref{fig:2}a in the main text). To that end, we deduce the following equation
\begin{equation}
C_\text{R}(\delta t)=C_\text{AS}(\delta t)+C_\text{Leak}=\alpha\cdot n_\text{m}(\delta t)+C_\text{Leak},\nonumber
\end{equation}
in which $\alpha$ is the constant of proportionality.

The long-term response of the mechanical mode to the initial blue pump pulse is shown in Methods Figure~\ref{fig:M3}a. It exhibits an exponential decay with a time constant of $T_\text{d}=34.4$~$\mu$s, which is interpreted as the mechanical damping time. The corresponding mechanical quality factor is then $Q=\omega_\text{m}\cdot T_\text{d}\approx 1.1\cdot10^6$.

In addition, we probe the short-term response of the mechanics within one microsecond after the blue pulse in more detail (Methods Figure~\ref{fig:M3}b). We observe an increase of $C_\text{R}$ with a time constant of 0.37~$\mu$s (fit to a simple exponential curve). This data reveals slow turn-on dynamics of pulse-induced heating, as previously studied in reference~\cite{Meenehan2015}. This time constant is even shorter than the decay of the cross-correlations (see Figure~\ref{fig:3}c in the main text), which we attribute to the increased thermal conductivity of silicon at higher temperatures, caused by absorption of increased optical pump energies.

\subsection{Characterization of the detection scheme}

\textit{Detection Efficiency}. We first calibrate the fiber-to-chip coupling efficiency ($\eta_\text{fc}$) by sending in light far off-resonant from the OMC cavity and then measure the reflected power ($\eta_\text{fc}=60.3$\% one-way). The device impedance ratio ($\eta_\text{c}$), i.e.\ the ratio of external coupling losses $\kappa_\text{ext}$ to total losses $\kappa_\text{c}$, is measured through the depth and the linewidth of the optical resonance, which we find to be $\eta_\text{c}=\kappa_\text{ext}/\kappa_\text{c}=0.5$. The detection efficiency of scattered photons for each detector ($\eta_\text{i}$; i=1,2) consists of $\eta_\text{fc}$, $\eta_\text{c}$, the total losses of the remaining detection paths ($\eta_\text{path,i}$), and the SNSPDs' quantum efficiencies ($\eta_\text{QE,i}$). To measure $\eta_\text{i}$, pulses with calibrated energy are sent off-resonantly to the OMC ($P_\text{in}$), and the reflected photons transmitted through the optical filters are detected by the SNSPDs ($P_\text{out}$). $P_\text{in}/P_\text{out}$ corresponds to $\eta_\text{fc}\cdot\eta_\text{fc}\cdot\eta_\text{path,i}\cdot\eta_\text{QE,i}$, which we measure to be 0.013 for SNSPD1 and 0.019 for SNSPD2. Therefore, we deduce $\eta_\text{i}=\eta_\text{c}\cdot\eta_\text{fc}\cdot\eta_\text{path,i}\cdot\eta_\text{QE,i}$ to be
\bea
\eta_1=1.1\%\nonumber\\
\eta_2=1.6\%.\nonumber
\eea
The detection efficiency of SNSPD1 (characterized quantum efficiency $\eta_\text{QE,1}=65$\%) is lower than SNSPD2 (characterized quantum efficiency $\eta_\text{QE,2}=90$\%), as we needed to reduce the bias current to prevent the detector from latching~\cite{Natarajan2012}. This latching is probably caused by a nearby heater of the dilution refrigerator. It also results in a slow drift in the quantum efficiency of SNSPD1. We note that the deduced $\eta_\text{path,i}$ come from the various optical elements in the beam path of the detection part and are in good agreement with their specified insertion losses.\\

\textit{Scattering rates and optomechanical coupling rate}. With the total detection efficiency of resonantly generated cavity photons, we can estimate the pair generation probability per write pulse (optical energy $E_\text{opt}$$\sim$40~fJ) to be $p$$\sim$3.0\%, including the effects of a finite starting temperature and leaked pump photons. The latter is calibrated by sending detuned optical pulses ($E_\text{opt}$$\sim$40~fJ, $\omega_\text{L}=\omega_\text{c}-\omega_\text{m}-2\pi\cdot200$~MHz) to the device. The generated optomechanical sidebands are now blocked by the filters and only leaked pump photons are detected. We measure a suppression of the pump pulse by 84~dB compared to an on- resonance transmission. Thus, approximately 1 out of 25 photons detected during the write pulse is a leaked pump photon. Knowing the scattering rate and the energy of the detuned pump pulse, we can determine the single-photon coupling rate of our OMC to be
\bea
g_0=\frac{\partial\omega_\text{c}}{\partial x}\sqrt{\frac{\hbar}{2m\omega_\text{m}}}=2\pi\cdot825~\text{kHz}.\nonumber
\eea
With this coupling rate, we can estimate the state-transfer efficiency of the red-detuned optical readout pulse of $E_\text{opt}=50$~fJ to be $\varepsilon_\text{R}=3.7$\%, where
\bea
\op{a}{opt,out}\approx\sqrt{1-\varepsilon_\text{R}}\op{a}{opt,in}+e^{i\phi}\sqrt{\varepsilon_\text{R}}\op{a}{mech,in}.\nonumber
\eea
Here, $\op{a}{opt,in(out)}$ are the annihilation operators of the temporal optical input (output) mode of the cavity resonance, $\op{a}{mech,in}$ the mechanical mode before the interaction, and $\phi$ an arbitrary but fixed phase between the inputs~\cite{Hofer2011}.

\subsection{Definition and properties of the second order correlation function}

We define the normalized second-order correlation function for two, not necessarily different modes $\alpha$ and $\beta$, and the respective annihilation operators $\op{a}{$\alpha$}$ and $\op{a}{$\beta$}$, to be (references~\cite{Kuzmich2003,Mandel1995} and references therein)
\bea
g_{\alpha\beta}^{(2)}=\frac{\langle :\opd{a}{$\alpha$}\op{a}{$\alpha$}\opd{a}{$\beta$}\op{a}{$\beta$}:\rangle}{\langle\opd{a}{$\alpha$}\op{a}{$\alpha$}\rangle\langle\opd{a}{$\beta$}\op{a}{$\beta$}\rangle},\nonumber
\eea
where $: \op{O}{} :$ denotes normal ordering of the operators. For the autocorrelation of the optical field (photons scattered by the write pulse), $\alpha=\beta=\text{o}$, for the mechanical field $\alpha=\beta=\text{m}$ and for the cross-correlation $\alpha=\text{o}$, $\beta=\text{m}$. By introducing effective modes $\gamma$, $\delta$ it can be seen that this correlation-function is independent of losses in the detection. Assuming the loss angles $\varphi_\alpha$ and $\varphi_\beta$ for detection of modes $\alpha$, $\beta$, we define the annihilation operators of the effectively detected modes $\gamma$, $\delta$
\bea
\op{a}{$\gamma/\delta$}=\cos(\varphi_{\alpha/\beta})\op{a}{$\alpha/\beta$}+\sin(\varphi_{\alpha/\beta})\op{l}{$\alpha/\beta$}\nonumber
\eea
by coupling the original modes $\alpha$, $\beta$ to modes $l_\alpha$ and $l_\beta$ represented by the annihilation operator \op{l}{$\alpha/\beta$}. As the detected modes $\gamma$, $\delta$ have frequencies in the optical domain, we can assume the in-coupled modes $l_\alpha$, $l_\beta$ to be in their respective ground state. Tracing over $l_\alpha$, $l_\beta$, we find that 
\bea
g_{\gamma\delta}^{(2)}=g_{\alpha\beta}^{(2)}\nonumber
\eea
i.e.\ the second order correlation function is independent of losses or, in the case of the mechanical mode, of "ineffective" partial state-transfer to the cavity mode. Thus, e.g.\ $g_{\text{mm}}^{(2)}$ is equivalent to the autocorrelation of the photons scattered by the read pulse.

For autocorrelation measurements, we use a Hanbury Brown and Twiss setup, by splitting the mode on a symmetric beamsplitter and sending it to a pair of detectors. We define the modes detected by the individual detectors $d_1$, $d_2$ with their annihilation operators
\bea
\op{a}{1/2}=\cos(\theta)\op{a}{$\alpha$}\pm\sin(\theta)\op{l}{d}\nonumber
\eea
with the splitting angle $\theta$ of the beam splitter and the annihilation operator \op{l}{d} of the second input of the beamsplitter. The input state can as before be approximated to be in its vacuum state. We find that the autocorrelation of mode $\alpha$ equals the cross-correlation of the two detectors:
\bea
g_{\alpha\alpha}^{(2)}=g_{12}^{(2)}.\nonumber
\eea
For a definition in terms of probabilities, see below.

\renewcommand{\figurename}{Methods Table}
\setcounter{figure}{0} 

\begin{figure*}[ht!]
\includegraphics[width=1.9\columnwidth]{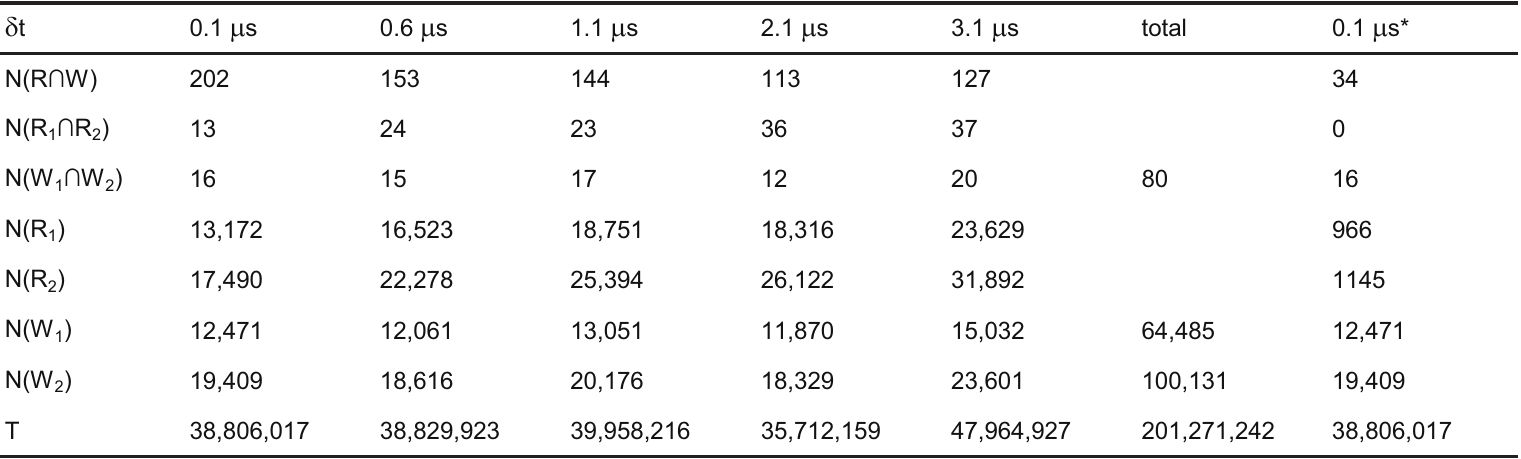}
\caption{\textbf{Counts of the cross-correlation measurements.} The row label 'N(event)' represents the number of counts for a certain event, e.g.\ detection of a photon during the measurement window of read pulse on detector 1/2 ($R_{1/2}$), or the coincidence of a detection event of a subsequent write and read pulse on either detector 1 or 2, $R\cap W=(R_1\cup R_2)\cap(W_1\cup W_2)$. $T$ denotes the total number of pulse pairs sent to the optomechanical device. For the calculation of the autocorrelation function of the read pulse, only counts from the delay setting $\delta t$ are used, as the delayed heating of the blue pulse (cf.\ Figure~\ref{fig:3}) influences the mechanical state. For the autocorrelation function of the write pulse, counts from all delay settings are summed, as the mechanical state is reinitialized by cryogenic cooling before measurement, independent of the delay $\delta t$. The numbers for this are summarized in the column labeled 'total'. The highest reported cross-correlation value was obtained by reducing the measurement window of the read pulse from 55~ns to 30~ns, with a delay of $\delta t=100$~ns between the write and the read pulse. The counts for this evaluation window are presented in the column marked with $^\ast$. The underlying dataset is the same as for the standard evaluation period of 55~ns, i.e.\ the first column.}
\label{table:M1}
\end{figure*} 

\subsection{Statistical Analysis}

Due to the low detection probability, the uncertainty in the estimation of the second-order correlation functions is completely dominated by the estimation of the coincidence rate $\langle :\opd{a}{$\alpha$}\op{a}{$\alpha$}\opd{a}{$\beta$}\op{a}{$\beta$}:\rangle$ of the two modes $\alpha$, $\beta$. As the absolute number of coincidences is low in some measurements, Gaussian statistics cannot be used for estimating uncertainties. Instead, we use the likelihood function based on the binomial distribution for estimating the probability $p$ of the underlying process, i.e.\ to obtain $N$ counts in $T$ tries
\bea
L(p,N,T)=\frac{1}{K}p^N(1-p)^{T-N}.\nonumber
\eea
The normalization $K$ is chosen such that $\int_0^1L(p,N,T)dp=1$. The upper and lower uncertainty $\sigma_+$ and $\sigma_-$ are chosen numerically, such that they cover a 68\% confidence interval around the maximum likelihood estimator $p_\text{ML}=N/T$, i.e.\  $\int_0^{p_\text{ML}-\sigma_-}L(p,N,T)dp=0.16$, $\int_{p_\text{ML}+\sigma_+}^1L(p,N,T)dp=0.16$.

For the classical bound of the cross-correlation, $g_{\text{cb}}^{(2)}=\sqrt{g_{\text{mm}}^{(2)}\cdot g_{\text{oo}}^{(2)}}$, the likelihood functions of the individual autocorrelations are convoluted. Due to their asymmetry, the maximum likelihood estimator of the classical bound is slightly lower than when using the individual maximum likelihood estimators $g_{\text{cb,ML}}^{(2)}\leq\sqrt{g_{\text{mm,ML}}^{(2)}\cdot g_{\text{oo,ML}}^{(2)}}$.

As estimators for the cross-correlation function, the probabilities $P$ of a coincidence- or single detection event during the read ($R$) and write pulse ($W$) were used, with $g_{\text{om}}^{(2)}=P(W\cap R)/P(R)P(W)$. This is valid for low event probabilities $P\ll1$. Autocorrelations were estimated by probabilities of coincidence- and single detection events on individual SNSPDs (1,2), $g_{\text{yy}}^{(2)}=P(X_1\cap X_2)/P(X_1)P(X_2)$ during the evaluation periods of the write (y = o, $X=W$) and read (y = m, $X=R$) pulse, respectively. The statistics of the cross-correlation measurements are summarized in Methods Table~\ref{table:M1}.

For the read pulse, only the first 55~ns of the pulse were evaluated (cf.\ Figure~\ref{fig:3}, main text). A further reduction of the evaluation period to $t_\text{eval}=30$~ns ($R^\ast$) has the advantage of reducing the influence of optical absorption of pump photons from the read pulse, while still obtaining solid statistics for the cross-correlation, $g_\text{om}^{(2)}(\Delta n=0,\delta t=100~\text{ns},t_\text{eval}=30~\text{ns})=19.6-2.8+3.9$. However, this reduction also results in a much lower state transfer efficiency $\varepsilon_\text{R}^\ast\approx0.1$\% (compared to $\varepsilon_\text{R}=3.7$\% above; see Methods section \textit{Characterization of the detection scheme}). As a consequence we cannot obtain independent statistics on the autocorrelation function of the read pulse, as the number of pulse sequences is too low to observe coincidences during the reduced read pulse $N(R_1^\ast\cap R_2^\ast)=0$. Thus, no independent classical bound $g_\text{cb}^\ast$ can be obtained for this case. As the measurement is identical to the one with longer evaluation window in the first column of Methods Table~\ref{table:M1}, it is reasonable to assume the same autocorrelation of the mechanical state and thus the same classical limit.

We note that slight differences in the polarization of the two input lasers and the optimal axis of the SNSPDs can lead to different detection rates of leaked pump photons between the read and the write pulse. While this does not influence the cross-correlation measurement, it is important to use the same laser source for the sideband asymmetry measurements.

\subsection{Interpretation of the cross-correlation measurements}

\textit{Classical bound}. The classical bound $g_\text{cb,ML}$ is found to be slightly above 2, the value expected for a thermal state of the mechanical system (cf.\ Figure~\ref{fig:3}c, main text). Although this increase of the autocorrelation is not significant in our measurements, a behavior like this would be expected in the case of mixed thermal states of different temperatures, caused e.g.\ by fluctuations in the absorbed power. Effects that usually decrease the measured autocorrelation function of a thermal state, such as dark counts of the detectors and instantaneous heating by the read pulse, do not play a major role in our experiment due to the choice of pulse parameters. In conclusion, the classical bound in the present experiment is slightly elevated compared to cross-correlation experiments in atomic physics or non-linear optics, where the classical bound is usually assumed to be~\cite{Sangouard2011} below 2.\\

\textit{Decay of cross-correlations due to delayed heating}. The cross-correlation can be interpreted as
\bea
g_\text{om}^{(2)}\sim\frac{\langle n_\text{m}\rangle_\text{h}}{\langle n_\text{m}\rangle}\nonumber
\eea
where $\langle n_\text{m}\rangle_\text{h}$ is the average number of mechanical excitations in the state heralded on a detection event of the write pulse (indicating the presence of an anti-Stokes scattered photon), and $\langle n_\text{m}\rangle$ the average number of unheralded events (essentially probing the thermal excitation of the system when $p\ll1$). In case of a delayed heating, the thermal occupation of the system is a function of the delay $\delta t$ after the write pulse $\langle n_\text{m,th}\rangle=\langle n_\text{m,th}\rangle(\delta t)$. Assuming our cross-correlation is dominated by the thermal occupation, we obtain for $\delta t\ll T_\text{d}$,
\bea
g_\text{om}^{(2)}(\delta t)\sim\frac{1+\langle n_\text{m,th}\rangle(\delta t)}{\langle n_\text{m,th}\rangle(\delta t)},\nonumber
\eea
which clearly decays in the case of substantial delayed heating as observed here (cf.\ Methods Figure~\ref{fig:M3}). Theoretical models of the complex thermodynamic non-equilibrium processes contain many device dependent parameters~\cite{Meenehan2015}, which will be subject of further studies.\\

\textit{Estimation of the heralded single-phonon fidelity}. In general, the toolbox of quantum optics provides unique means for quantum state control of various systems~\cite{Zoller2004}. As an example we discuss the application of single-photon detection for the heralded generation of single-phonon Fock states of our mechanical resonator~\cite{Hofer2011,Vanner2013,Galland2014}. To estimate the fidelity of the single-phonon state directly after heralding on the detection of a resonant photon generated by the write pulse, we need to know all contributions to the diagonal of the density matrix, which are not a single phonon. These contributions can either be higher excitations by thermal contribution, multi-pair generation, or vacuum states by false positive heralding events. Higher excitations can be estimated by the auto-correlation function of the heralded state, which is related to the cross-correlations function~\cite{Zhao2009}. As the target is to estimate the state immediately after heralding it, we reduce the evaluation window of the read pulse as much as possible, while maintaining reasonable statistics on the cross-correlation (cf.\ Methods Table~\ref{table:M1}). From measured $g_\text{om}^{(2)}(\Delta n=0,\delta t=100~\text{ns},t_\text{eval}=30~\text{ns})=19.6-2.8+3.9$, we infer an auto-correlation function for the heralded mechanical state of $g_\text{mm,heralded}^{(2)}\approx0.22\pm0.04$, which approximately relates to the ratio of probabilities of higher excitations $p_{n>1}$ to single phonon excitations $p_{n=1}$, $2\cdot p_{n>1}\approx g_\text{mm,heralded}^{(2)}\cdot p_{n=1}^2$. In the meantime, the main contribution for non-zero $p_{n=0}$ (i.e.\ the probability of the heralded mechanical state being the ground state) is false positive heralding events, i.e.\ dark counts and leaked pump photons. With the known ratio of true positive to false positive heralding events (cf.\ section \textit{Characterization of the detection scheme}), we obtain an estimate of $p_{n=0}\sim1/25$. With these conservative estimates, we obtain a heralded Fock-state fidelity of $p_{n=1}=87.7\pm1.2$\% on the basis of the standard system Hamiltonian of the optomechanical device~\cite{Hofer2011}.

\end{document}